\documentclass[10pt]{IEEEtran}

\newcommand{\ie}{{\it i.e.}}

\newcommand{\argmax}{\mathop{\rm argmax}}
\newcommand{\argmin}{\mathop{\rm argmin}}

\newcommand{\BA}{\begin{array}}
\newcommand{\EA}{\end{array}}
\newcommand{\BEAS}{\begin{eqnarray*}}
\newcommand{\EEAS}{\end{eqnarray*}}
\newcommand{\BEA}{\begin{eqnarray}}
\newcommand{\EEA}{\end{eqnarray}}
\newcommand{\BEQ}{\begin{equation}}
\newcommand{\EEQ}{\end{equation}}
\newcommand{\BIT}{\begin{itemize}}
\newcommand{\EIT}{\end{itemize}}
\newcommand{\BNUM}{\begin{enumerate}}
\newcommand{\ENUM}{\end{enumerate}}
\newcommand{\SNR}{\mathop{\mathsf{SNR}}}

\usepackage{citesort}
\usepackage[dvips]{graphicx}
\usepackage[usenames,dvipsnames]{color}
\usepackage{amssymb,amsmath}
\usepackage{multirow}

\usepackage{amsmath}   % From the American Mathematical Society
\usepackage{amsmath}   % From the American Mathematical Society
\begin{document}
\title{Code Diversity in Multiple Antenna Wireless Communication}

\author{Yiyue Wu and Robert Calderbank\\ Department of Electrical Engineering, Princeton University,\\ Princeton NJ 08544 USA. Email: \{yiyuewu,calderbk\}@princeton.edu %Department of Electrical Engineering, Princeton University\\
}

\maketitle

\begin{abstract}
The standard approach to the design of individual space-time codes
is based on optimizing diversity and coding gains. This geometric
approach leads to remarkable examples, such as perfect space-time
block codes \cite{Oggier}, for which the complexity of Maximum
Likelihood (ML) decoding is considerable. \emph{Code diversity} is
an alternative and complementary approach where a small number of
feedback bits are used to select from a family of space-time codes.
Different codes lead to different induced channels at the receiver,
where Channel State Information (CSI) is used to instruct the
transmitter how to choose the code. This method of feedback provides
gains associated with beamforming while minimizing the number of
feedback bits. Thus code diversity can be viewed as the integration
of space-time coding with a fixed set of beams. It complements the standard approach to code design
by taking advantage of different (possibly equivalent) realizations
of a particular code design. Feedback can be combined with
sub-optimal low complexity decoding of the component codes to match
ML decoding performance of any individual code in the family. It can
also be combined with ML decoding of the component codes to improve
performance beyond ML decoding performance of any individual code.
One method of implementing code diversity is the use of feedback to
adapt the phase of a transmitted signal. Phase adaptation with the
$4\times4$ Quasi-Orthogonal Space-Time Block Code (QOSTBC) is shown
to be almost information lossless; that is, this form of space-time
coding does not reduce the capacity of the underlying multiple
antenna wireless channel. Code diversity can also be used to improve
performance of multi-user detection by reducing interference between
users. Phase adaptation with two Alamouti users makes it possible
for the Zero Forcing (ZF) or decorrelating detector to match the
performance of ML joint detection. Code diversity implemented by
selecting from equivalent variants is used to improve ML decoding
performance of the Golden code. This paper introduces a family of
full rate circulant codes which can be linearly decoded by fourier
decomposition of circulant matrices within the code diversity
framework. A $3\times3$ circulant code is shown to outperform the
Alamouti code at the same transmission rate.

%Easy implementation together with significant improvement on
%performance and large reduction on decoding complexity makes it
%indeed practical.
\end{abstract}
\vspace{0.1cm}
\begin{IEEEkeywords}
Multiple antennas, wireless communication, space-time codes,
information lossless, diversity
\end{IEEEkeywords}

\section{Introduction}
Space-time codes improve the reliability of communication over
fading channels by correlating signals across different transmit
antennas. Two design criteria were proposed by Tarokh {\it et al.}
\cite{Tarokh}. The first is the rank criterion, which is to maximize
the minimum rank ($r$) of the difference $X_{1}-X_{2}$ over all
pairs of distinct codewords $X_{1}, X_{2}$; the second is the
determinant criterion, which is to maximize the minimum product
($\gamma$) of all nonzero singular values of the difference
$X_{1}-X_{2}$ over all pairs of distinct codewords $X_{1}, X_{2}$.
The quantity $r$ is the diversity gain and the quantity $\gamma$
determines the coding gain. Diversity and coding gains determine the
performance of space-time codes in the high $\SNR$ regime.

The ML decoder provides optimal decoding performance, but complexity
typically becomes prohibitive as the constellation size increases.
Sphere decoding \cite{Jalden} aims to reduce decoding complexity by
restricting to a search over lattice points that lie in a sphere
about an initial estimate. However when the channel matrix is close
to singular (as will be the case on Line of Sight channels), the
preprocessing stage of sphere decoding yields a plane of
possibilities rather than a single initial estimate. Jald\'{e}n {\it
et al.} \cite{Jalden} showed that overall expected complexity is
exponential in transmission alphabet size and strongly dependent on
signal to noise ratio ($\SNR$). Decoders such as the zero forcing
decoder have low complexity but are suboptimal in terms of decoding
performance.

Tan and Calderbank \cite{tan} introduced code diversity, a form of
adaptive modulation where feedback is used to instruct the
transmitter how to choose from a plurality of codes. In the case of
QOSTBC, they showed that ML decoding of a particular code may be
inferior (in terms of both error probability and decoding
complexity) to a combination of adaptive modulation and suboptimal
decoding. Bonnet {\it et al.} \cite{Bonnet} examined the value of
closed loop feedback in the context of the DSTTD and DABBA
space-time block codes. Feedback is also examined by C. Lin {\it et
al.} \cite{Chelin} in the context of beamforming.

This paper provides a general information theoretic framework for
the analysis of code diversity. We show that it is possible to
quantify and sometimes eliminate the gap between the error
performance of low complexity zero forcing decoders and that of the
optimal decoder. Code diversity increases the mutual information
between the transmitted and received signals and it is possible to
approach the information lossless property for certain codes, such
as the QOSTBC \cite{Jafark1,Jafark2,Su,tan}. We introduce methods of
inducing code diversity, such as adapting the phases of channel
gains via a small number of feedback bits. Simulation shows that a
small number of feedback bits are sufficient to obtain close to
optimal decoding performance even with the simplest zero forcing
decoder.

Feedback has also been used for opportunistic beamforming by
Viswanath {\it et al.} \cite{Pramod} to optimize the total
throughput for multi-user systems by modifying the phases of channel
gains and the transmitted power at each transmit antenna. These
authors show that opportunistic beamforming can approach the
performance of perfect beamforming for systems with a large number
of users. However it requires centralized processing and does not
work as well for systems with fewer users. The focus of this paper
is not sum capacity, rather it is improving the performance of
individual links or joint detection of small number of users. Our
use of phase adaptation avoids centralized processing and we
demonstrate that our code diversity scheme brings significant
improvement to the performance of general space-time codes  and
large reduction in decoding complexity.

We also introduce a family of full rate code designs based on
circulant matrices within the code diversity framework. A universal
linear decoder for this family of codes is also proposed based on
fourier decomposition of circulant matrices. A $3\times3$ circulant
design is analyzed and shown to outperform the Alamouti code for a
given transmission rate.

The paper is organized as follows. Section \ref{sysmod} presents the
system model, Section \ref{FLFRC} discusses fundamental limits on
full rate space-time codes. Our code diversity scheme is introduced
and analyzed in Section \ref{Phaseadaptation}, applications are
explored in Section \ref{application},  and the new family of
circulant codes is introduced in Section  \ref{CCD}. Section
\ref{conclusion} provides conclusions.

\section{System Model}
\label{sysmod}A space-time block codeword $X$ is represented as an
$M\times T$ matrix, where the rows are indexed by transmit antennas
and the columns are indexed by time slots. If $L$ distinct symbols
are transmitted during one frame of $T$ symbol periods, then the
transmission rate for the space time block code is $\frac{L}{T}$. In
a system with $M$ transmit antennas and $N$ receive antennas, the
received signal $\tilde{r}\in \mathcal{C}^{N\times T}$ is given by
\begin{equation}
\label{model0} \tilde{r}=\sqrt{\frac{E_{s}}{MN_{0}}}HX+\tilde{n}
\end{equation}
where $H\in\mathcal{C}^{N\times M}$ is the channel matrix with entry
$h_{ji}$ representing the channel gain between the $i^{th}$ transmit
antenna and the $j^{th}$ receive antenna; $X\in \mathbb{C}^{M\times
T}$ is a codeword with entries in the constellation $\mathbb{C}$;
$\tilde{n}\in\mathcal{C}^{N\times T}$ is the normalized additive
noise with entries as complex Gaussian with zero mean and unit
variance, \ie\;\ $\mathcal{CN}(0, 1)$; $N_0$ is the actual noise
variance; and $E_{s}$ is average transmitted signal power.

An important feature of many algebraic constructions of space-time
codes is \emph{transference} of structure (correlation) at the
transmitter to structure at the receiver. Transference means, we can
rewrite (\ref{model0}) as:
\begin{equation}
\label{model1} r=\sqrt{\frac{E_{s}}{MN_{0}}}\mathcal{H}c+n
\end{equation}
where $r\in \mathcal{C}^{NT\times 1};\;\
\mathcal{H}\in\mathcal{C}^{NT\times L}$; $c\in\mathbb{C}^{L\times
1}$ is the transmitted signal; and $n\in\mathcal{C}^{NT\times 1}$.
Since $\SNR = \frac{E_{s}}{N_{0}}$, equation (\ref{model1}) can also
be written as $ r=\sqrt{\frac{\SNR}{M}}\mathcal{H}c+n$.

Equation (\ref{model1}) captures the perspective of the receiver,
where the induced channel is assumed to be known, and the problem is
to estimate the transmitted signals. Transference of structure from
transmitter to receiver is possible for (real-valued) linear
dispersion codes \cite{Hassibi1}, the Golden code
\cite{Belfiore,Yao,Dayal}, the Silver code \cite{HKT,HT,HTW,TA,TK,Paredes} and many
more.

We assume that the wireless channel is quasi-static, that is it
remains constant over a frame and changes independently from one
frame to the next.

\subsection{Channel Capacity}
Given perfect CSI at the receiver, the
channel capacity of the system with channel matrix $H$ is
\cite{Telatar}:

\begin{equation}
\label{capacity2} C_{0}(M,N)=\mathbf{E}_{H}\log\det\left
(I_{N}+\frac{E_{s}}{MN_{0}}HH^{\dag}\right )
\end{equation}
where, $\mathbf{E}_{H}$ is the expectation with respect to $H$.

For a specific space time block code with an induced channel matrix
$\mathcal{H}$, the achievable maximum capacity is \cite{Damen1}:
\begin{equation}
\label{capacity4}
C(M,N)=\frac{1}{T}\mathbf{E}_{\mathcal{H}}\log\det\left
(I_{NT}+\frac{E_{s}}{MN_{0}}\mathcal{H}\mathcal{H}^{\dag}\right )
\end{equation}
The space time code is said to be {\it information lossless} if
$C(M,N)=C_{0}(M,N)$.

For simplicity, we focuse on wireless systems with multiple transmit
antennas and a single receive antenna (MISO).

\section{Fundamental Limits on Full Rate Codes}
\label{FLFRC}
We consider full rate square designs $X$ that are linear and that
employ QAM signaling. The average power constraint is expressed in
terms of the Frobenius norm by $\|X\|_F^2 \leq P$. Given linearity,
the standard design criterion reduces to maximizing the minimum
determinant among all possible codewords.

\subsection{Optimality of Orthogonal Designs}

If $X$ is an $M\times M$ block space-time code with columns $C_i$,
$i=1,\cdots, M$, then the hadamard inequality states

\begin{equation}
\label{hadamardE}
|\det{X}|\leq  \prod_{i=1}^{M}||C_i||
\end{equation}

Equality holds for orthogonal designs, that is when $C_i C_j^{\dag}=0$ for all $i\neq j$.

G. Bresler and B. Hajek \cite{Hajek} proved that
orthogonal designs also maximize the mutual information subject to
an average power constraint on each symbol. For example, the Alamouti code is shown to be
information lossless (namely maximizing mutual information with
certain power constraints) for $2\times 1$ systems (see Hassibi and
Hochwald \cite{Hassibi1}). It is easily seen that any $M\times M$
orthogonal design is information lossless for $M\times 1$ systems.

\subsection{Existence of Orthogonal Designs}

Wolfe's theorem \cite{Wolfe} on amicable pairs of real orthogonal designs
provides fundamental limits on the achievable rates of space-time block codes
that employ complex signaling (see \cite{Calderbank,Tarokh1,Liang}). If $M=2^h M_0$,
where $M_0$ is odd, then the maximum rate of a complex orthogonal design is
$\frac{h+1}{M}$. When $M=2$, the Alamouti code is optimal, and when $M=4$, the
orthogonal space-time code
\begin{equation}
\label{OSTBC}  X(x_1,x_2,x_3)=\left ( \begin{array}{cccc} x_{1} & x_{2} & x_{3} & 0\\
-x^*_2 & x^*_1 & 0 & x_3\\
-x_3^* & 0 & x_1^* & -x_2\\
0 & -x^*_{3} & x_{2}^* & x_1 \end{array} \right)
\end{equation}
achieves the optimal rate.

Complex orthogonal designs maximize both the minimum determinant and
mutual information, but fail to achieve full rate for more than two transmit antennas. Code diversity
offers a way forward.

\section{Code Diversity Through Phase Adaptation}
\label{Phaseadaptation} The mechanism of phase adaptation is a simple and effective approach to code diversity.
Other possible mechanisms are examined in section \ref{application}.

\subsection{Implementation}
Equation (\ref{model1}) transfers structure at the transmitter (code
structure) to structure at the receiver (induced channel structure).
If there is no noise in (\ref{model1}), the ML estimate for the
transmitted symbols $c$ is unique provided the column rank $\mathcal{R}(\mathcal{H})$ of channel matrix
$\mathcal{H}$ is full, \ie\;\
$\mathcal{R}(\mathcal{H})=L$. This may not always be satisfied, and when
$\mathcal{H}$ is close to rank deficient, decoding becomes unstable.
We propose the following algorithm to combat rank deficiency of the
channel matrix $\mathcal{H}$.

We allow the receiver to send back feedback to
modify phases of channel gains as follows
$$h_{ij}\rightarrow h_{ij}\times e^{i\frac{2\pi k_{ij}}{K}}$$
where  $k_{ij}\in\{1,2...,K\}$ is the feedback information, $K$
determines the number of feedback bits, and
$\mathcal{H}^{\dag}\mathcal{H}$ changes with different phase
adaptations. Our algorithm only modifies phases of a small subset of
the channel gains. For example, only the phase of one channel gain needs
to be modified for systems with four transmit antennas and a single
receive antenna.

Let $\{h_{ij}\}$ denote the set where the phases of channel gains
$h_{ij}$ are modified and let $\{\hat{k}_{ij}\}$ denote the feedback
selection information. Then the receiver selects $\{\hat{k}_{ij}\}$
which

\begin{itemize}
\item firstly maximizes $\mathcal{R}(\mathcal{H}^{\dag}\mathcal{H})$
(Let $d_{\max}=\max\mathcal{R}(\mathcal{H}^{\dag}\mathcal{H})$ )
\item and secondly maximizes $\left(\prod_{i=1}^{d_{\max}}\omega_{i} \right)$, where $\omega_{i}$ are the nonzero eigenvalues of
$\mathcal{H}^{\dagger}\mathcal{H}$.
\end{itemize}

If $d_{\max}=L$, \ie\;\ $\mathcal{H}^{\dag}\mathcal{H}$ has full
rank, then the receiver selects $\{\hat{k}_{ij}\}$ as
\begin{equation}
\label{stra}
\{\hat{k}_{ij}\}=\argmax_{k_{ij}\in\{1,2...,K\}}det(\mathcal{H}^{\dag}\mathcal{H})|_{\{h_{ij}\}\rightarrow
\{h_{ij\times}e^{i\theta_{ij}}\}}
\end{equation}

\begin{figure}[h!]
\begin{center}
\resizebox{6cm}{!}{\includegraphics{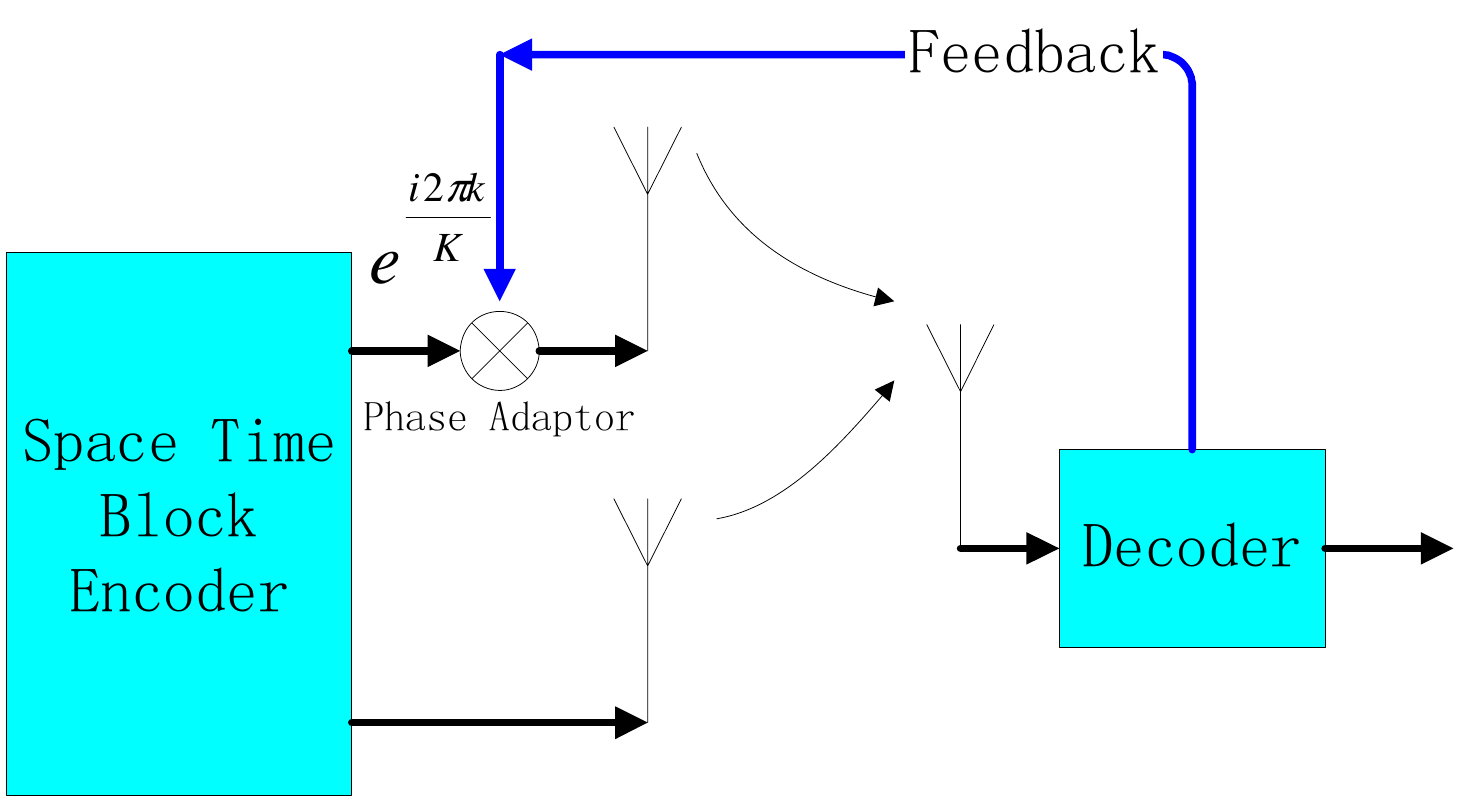}} \caption{Code
diversity implemented through phase adaptation for a system with two
transmit antennas and one receive antenna.} \label{fig:graph3}
\end{center}
\end{figure}

The structure in Fig. \ref{fig:graph3} illustrates our strategy for
the system with two transmit antennas and one receive antenna, where
feedback is used to adapt the phase of one channel gain.

\textbf{Remark}: In the above algorithm, code diversity is induced
by different channel phase adaptations. Via feedback, the receiver
can select the best code with a particular phase adaptation. Phase
adaptation is also employed in \cite{Bonnet}. Code diversity may
also be induced in other ways.

In contrast with opportunistic beamforming, feedback is employed to
increase reliability and achievable capacity and to enable low
complexity decoders for each transmission. We now show that the above algorithm
optimizes error performance for general space time codes and
also maximizes achievable capacity $C(M,N)$ given in equation
(\ref{capacity4}).

\subsection{Analysis and Discussion}
\label{AnalysisE}

\subsubsection{Analysis of Error Probability}
For simplicity we restrict to systems with multiple transmit antennas and a
single receive antenna ($N=1$). Assuming CSI at the receiver, the
conditional probability of deciding $X'$ when transmitting $X$ is:
\begin{equation}
\label{Ana1} P(X\rightarrow X'|H)\leq
exp(-\frac{E_{s}}{4MN_{0}}\parallel H(X-X')
\parallel_{F}^{2})
\end{equation}
where, $\|\cdot\|_{F}$ denotes the matrix Frobenius norm (see
\cite{Tarokh}). Given (\ref{model1}), we rewrite (\ref{Ana1}) as
\begin{equation}
P(c\rightarrow c'|\mathcal{H})\leq
exp(-\frac{E_{s}}{4MN_{0}}\parallel \mathcal{H}(c-c')
\parallel_{F}^{2})
\end{equation}

Then, the conditional average pairwise error probability is
\begin{equation}
\begin{array}{ll}
\label{Ana2} P_{e|\mathcal{H}}=P(\hat{c}\neq c|\mathcal{H})\\
\leq \mathbf{E}_{D}\{exp(-\frac{E_{s}}{4MN_{0}}\parallel
\mathcal{H}D
\parallel_{F}^{2})\}\\
=\mathbf{E}_{D}\{exp(-\frac{E_{s}}{4MN_{0}}D^{\dagger}\mathcal{H}^{\dagger}\mathcal{H}D)\}
\end{array}
\end{equation}
where $D\triangleq c-c'$.

To simplify this upper bound, we assume that $D\sim
\mathcal{CN}(0,I_{L})$ and use the fact that
\begin{equation}
\mathbf{E}_{s}\{exp(-s^{\dagger}\Psi
s)\}=\frac{exp(-\mu_{s}^{\dagger}\Psi[I+\Sigma\Psi]^{-1}\mu_{s})}{det(I+\Sigma\Psi)}
\end{equation}
for $s\sim\mathcal{CN}(\mu_{s},\Sigma)$ and a Hermitian matrix $\Psi$.

We have
\begin{equation}
\begin{array}{ll}
\label{Ana3} P_{e|\mathcal{H}}\leq
\frac{1}{det(I_{L}+\frac{E_{s}}{4MN_{0}}\mathcal{H}^{\dagger}\mathcal{H})}
=\frac{1}{\prod_{i=1}^{d}(1+\frac{E_{s}}{4MN_{0}}\omega_{i})}\\
\approx\left (\prod_{i=1}^{d}\omega_{i} \right)^{-1}\left
(\frac{E_{s}}{4MN_{0}} \right)^{-d}.
\end{array}
\end{equation}
where $d=\mathcal{R}(\mathcal{H}^{\dagger}\mathcal{H})$ and
$\omega_{i},i=1,\cdots,d$ are the nonzero eigenvalues of
$\mathcal{H}^{\dagger}\mathcal{H}$. The diversity gain $d$ and
coding gain $\left(\prod_{i=1}^{d}\omega_{i} \right)$ determine
space-time code performance.

Code diversity optimizes space-time code performance by maximizing
the diversity and coding gain.

Since
$\mathcal{R}(\mathcal{H}^{\dagger}\mathcal{H})=\mathcal{R}(\mathcal{H})\leq
\min(T,L)$, we have $d_{\max}=\min(T,L)$. For typical space-time block codes, $\min(T,L)\geq
M$, and for such space-time block code, we have
\begin{equation}
\label{Ana5} P_{e|\mathcal{H}}\leq  \left (\prod_{i=1}^{M}\omega_{i}
\right)^{-1}\left (\frac{E_{s}}{4MN_{0}} \right)^{-M}
\end{equation}

The above analysis shows that the error performance can be optimized
by our code diversity scheme. Since $\mathcal{H}$ is adjusted to
avoid rank deficiency, it is possible to employ low complexity zero-forcing
decoding schemes.

\subsubsection{Analysis of Achievable Capacity}
The achievable capacity $C(M,N)$ in equation (\ref{capacity4}) can
be further written as
\begin{equation}
\label{capacityAppro}
\begin{array}{ll}
C(M,N)=\frac{1}{T}\mathbf{E}_{\mathcal{H}}\log\prod_{i=1}^{d}\left(1+\frac{E_{s}}{MN_{0}}\omega_{i}\right)\\
\approx\frac{1}{T}\mathbf{E}_{\mathcal{H}}\log\left
(\prod_{i=1}^{d}\omega_{i} \right)\left (\frac{E_{s}}{MN_{0}}
\right)^{d}
\end{array}
\end{equation}

This shows that $\left (\prod_{i=1}^{d}\omega_{i} \right)$ and $d$
determine achievable capacity; the aim of code diversity is to
maximize these quantities.

\subsubsection{Discussion of Decoding Schemes} Given perfect
CSI at the receiver, ML decoding achieves the optimal decoding
performance at the receiver. For $2^{m}$-QAM signaling using the
constellation $\mathbb{C}_m$, the ML decoding scheme
is formulated as
\begin{equation}
\label{MLdec} \hat{c}=\argmin_{c\in
\mathbb{C}_{m}^L}\|r-\sqrt{\frac{E_{s}}{MN_{0}}}\mathcal{H}c\|^{2}
\end{equation}
Code diversity improves the performance/complexity tradeoff in decoding algorithms.

As the constellation size increases and the dimension of STBC goes
up, ML decoding becomes computationally prohibitive. To reduce
decoding complexity, we can adopt zero forcing decoding. However,
near singularity of the channel matrix $\mathcal{H}$ in
(\ref{model1}) reduces the effectiveness of these low-complexity
schemes. Our proposed method increases the rank of the matrix $\mathcal{H}$ and avoids noise amplification. For the general receive
structure (\ref{model1}), the simplest zero forcing decoding scheme
is formulated as
\begin{equation}
\label{zfdec} \hat{c}=\argmin_{c\in \mathbb{C}_{m}^L}
\|\mathcal{H}^{-1}r-\sqrt{\frac{E_{s}}{MN_{0}}}c\|^{2}
\end{equation}
which decouples into $L$ subproblems each with linear decoding
complexity. The value of code diversity is that the simplest zero
forcing scheme is often good enough to obtain close to the performance of ML
decoding of one of the constituent codes.

\section{Applications}
\label{application}
\subsection{Quasi-Orthogonal Space Time Block Codes}
In a system with four transmit antennas and one receive antenna, we
consider the following quasi-orthogonal space time block code
proposed by Jafarkhani \cite{Jafark1}:
\begin{equation}
\label{jafarkQOSTBC}  X=\left ( \begin{array}{cccc}
x_{1} & x_{2} & x_{3} & x_{4}\\
-x_{2}^{*} & x_{1}^{*} & -x_{4}^{*} & x_{3}^{*}\\
-x_{3}^{*} & -x_{4}^{*} & x_{1}^{*} & x_{2}^{*}\\
x_{4} & -x_{3} & -x_{2} & x_{1} \end{array} \right)
\end{equation}

As shown in (\ref{Ana5}), the conditional average pairwise error
probability for this QOSTBC under our code diversity scheme is upper
bounded as
\begin{equation}
P_{e|\mathcal{H}}\leq \left (\prod_{i=1}^{4}\omega_{i}
\right)^{-1}\left (\frac{E_{s}}{16N_{0}} \right)^{-4}
\end{equation}

We now verify that our code diversity scheme optimizes the error
performance and does not significantly reduce the capacity of the
underlying multiple antenna wireless channel.

\subsubsection{Algorithm Implementation}
With one receive antenna, the received signal can be written as
\begin{equation}
\label{QOSTBC}  r=\left ( \begin{array}{cccc}
h_{1} & h_{2} & h_{3} & h_{4}\\
-h_{2}^{*} & h_{1}^* & -h_{4}^{*} & h_{3}^*\\
-h_{3}^{*} & -h_{4}^{*} & h_{1}^{*} & h_{2}^{*}\\
h_{4} & -h_{3} & -h_{2} & h_{1} \end{array} \right)
\left (\begin{array}{c} x_{1}\\
x_{2}\\
x_{3}\\
x_{4} \end{array} \right) +\left (\begin{array}{c} n_{1}\\
n_{2}\\
n_{3}\\
n_{4} \end{array} \right)
\end{equation}
where $h_{i}$ is the channel gain between the $i^{th}$ transmit
antenna and the single receive antenna. For simplicity, we denote it
as
\begin{equation}
\label{QOSTBC}  r=\mathcal{H}c+n.
\end{equation}

Then we have
\begin{equation}
\mathcal{H}^{\dagger}\mathcal{H}=\mathcal{H}\mathcal{H}^{\dagger}=\left
(
\begin{array}{cccc}
a & 0 & 0 & b\\
0 & a & -b & 0\\
0 & -b & a & 0\\
b & 0 & 0 & a \end{array} \right)
\end{equation}
where $a=\sum_{i=1}^{4}|h_{i}|^{2}$ and\\
$b=h_{1}h_{4}^{*}+h_{1}^{*}h_{4}-h_{2}h_{3}^{*}-h_{2}^{*}h_{3}=2\Re(h_{1}h_{4}^{*}-h_{2}h_{3}^{*})$.
\\Hence $det(\mathcal{H}^{\dagger}\mathcal{H})=a^{4}-b^{4}$.

Notice $$a-b=|h_{1}-h_{4}|^{2}+|h_{2}+h_{3}|^{2}\geq 0$$
$$a+b=|h_{1}+h_{4}|^{2}+|h_{2}-h_{3}|^{2}\geq 0$$ therefore $det(\mathcal{H}^{\dagger}\mathcal{H})\geq 0$.

We notice $a$ is unchanged by rotation. Given phase adaptation on
the phase of $h_{1}$, the feedback algorithm is
$$\hat{k}= \argmax_{k\in\{1,2...,K\}}
det(\mathcal{H}^{\dagger}\mathcal{H})|_{h_{1}\rightarrow
h_{1}e^{i\frac{2\pi k}{K}}}$$
\begin{equation}
\begin{array}{ll}
\label{QOSTBCalg1} =\argmin_{k\in\{1,2...,K\}} |b|_{h_{1}\rightarrow
h_{1}e^{i\frac{2\pi k}{K}}}
\end{array}
\end{equation}

The optimum phase adaptation for $h_{1}$ is $\hat{\theta}=\angle
\frac{h_{2}h_{3}^{*}}{h_{1}h_{4}^{*}}$ ($\hat{\theta}$ is the phase
of $\frac{h_{2}h_{3}^{*}}{h_{1}h_{4}^{*}}$). Therefore, the
algorithm described in (\ref{QOSTBCalg1}) can be expressed as
\begin{equation}
\label{QOSTBCalg2} \hat{k}= \argmin_{k\in\{1,2...,K\}}|\frac{2\pi
k}{K}-\hat{\theta}|
\end{equation}

\begin{figure}[h!]
\begin{center}
\resizebox{9cm}{!}{\includegraphics{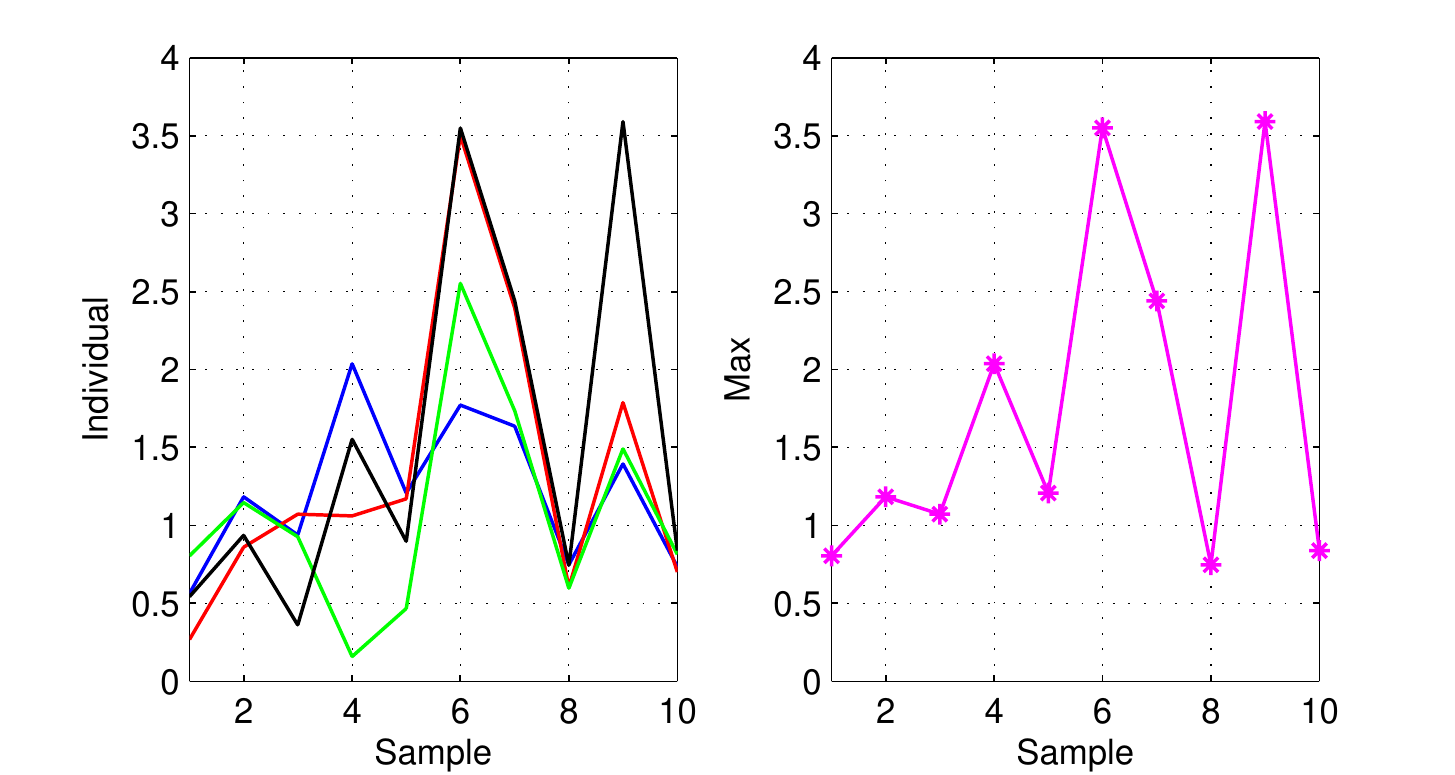}} \caption{Illustration of the value of code
diversity showing the improvement that results from choosing the best of four induced channels.}
\label{fig:QSTBCdet}
\end{center}
\end{figure}

This algorithm is illustrated in Fig. \ref{fig:QSTBCdet} where the
best code with the largest
$\sqrt{det(\mathcal{H}^{\dagger}\mathcal{H})}$ is picked among four
codes induced by phase adaptation ($K=4$).

With sufficient feedback, $|b|\approx 0$, hence
$\mathcal{H}^{\dagger}\mathcal{H}\approx\sum_{i=1}^{4}|h_{i}|^{2}\times
I_{4}$; therefore, the upper bound for the conditional average
pairwise error probability under our code diversity scheme is
\begin{equation}
P_{e|\mathcal{H}}\leq \left (\prod_{i=1}^{4}\omega_{i}
\right)^{-1}\left (\frac{E_{s}}{16N_{0}}
\right)^{-4}=(\sum_{i=1}^{4}|h_{i}|^{2})^{-4}\left
(\frac{E_{s}}{16N_{0}} \right)^{-4}
\end{equation}
Since $\mathcal{H}^{\dagger}\mathcal{H}$ is adjusted to have full
rank and maximum determinant, the error performance is optimized.

The maximum capacity for the system with four transmit antennas and
one receive antenna is
\begin{equation}
\begin{array}{ll}
\label{capacity2} C_{0}(4,1)=\mathbf{E}_{H}\log\det\left
(I_{1}+\frac{E_{s}}{4N_{0}}HH^{\dag}\right )\\
=\mathbf{E}_{H}\log\det\left
(1+\frac{E_{s}}{4N_{0}}\sum_{i=1}^{4}|h_{i}|^{2}\right )
\end{array}
\end{equation}

The achievable capacity for the QOSTBC under the feedback algorithm
can be formulated as
\begin{equation}
\label{capacityQOSTBC}
\begin{array}{ll}
C(4,1)=\frac{1}{4}\mathbf{E}_{\mathcal{H}}\log\det\left
(I_{4}+\frac{E_{s}}{4N_{0}}\mathcal{H}\mathcal{H}^{\dag}\right )\\
\approx\frac{1}{4}\mathbf{E}_{\mathcal{H}}\log\det\left
(I_{4}\times(1+\frac{E_{s}}{4N_{0}}\sum_{i=1}^{4}|h_{i}|^{2})\right
)\\
=\mathbf{E}_{\mathcal{H}}\log\det\left
(1+\frac{E_{s}}{4N_{0}}\sum_{i=1}^{4}|h_{i}|^{2}\right )
\end{array}
\end{equation}

Therefore, $C(4,1)\approx C_{0}(4,1)$, \ie\;\ maximum capacity is
closely approached and QOSTBC with code diversity is approximately
information lossless.

\subsubsection{Simulation}
The CSI is assumed to be known at the receiver and each channel is
modeled as complex Gaussian with zero mean and unit variance. The
noise is complex Gaussian with zero mean and unit variance. Then the
$\SNR$ is defined as
$$\SNR=\frac{E_{s}}{N_{0}}\frac{\|H\|_{F}^{2}}{4}$$
Where $E_{s}$ is the average transmit symbol power.

\begin{figure}[h!]
\begin{center}
\resizebox{7cm}{!}{\includegraphics{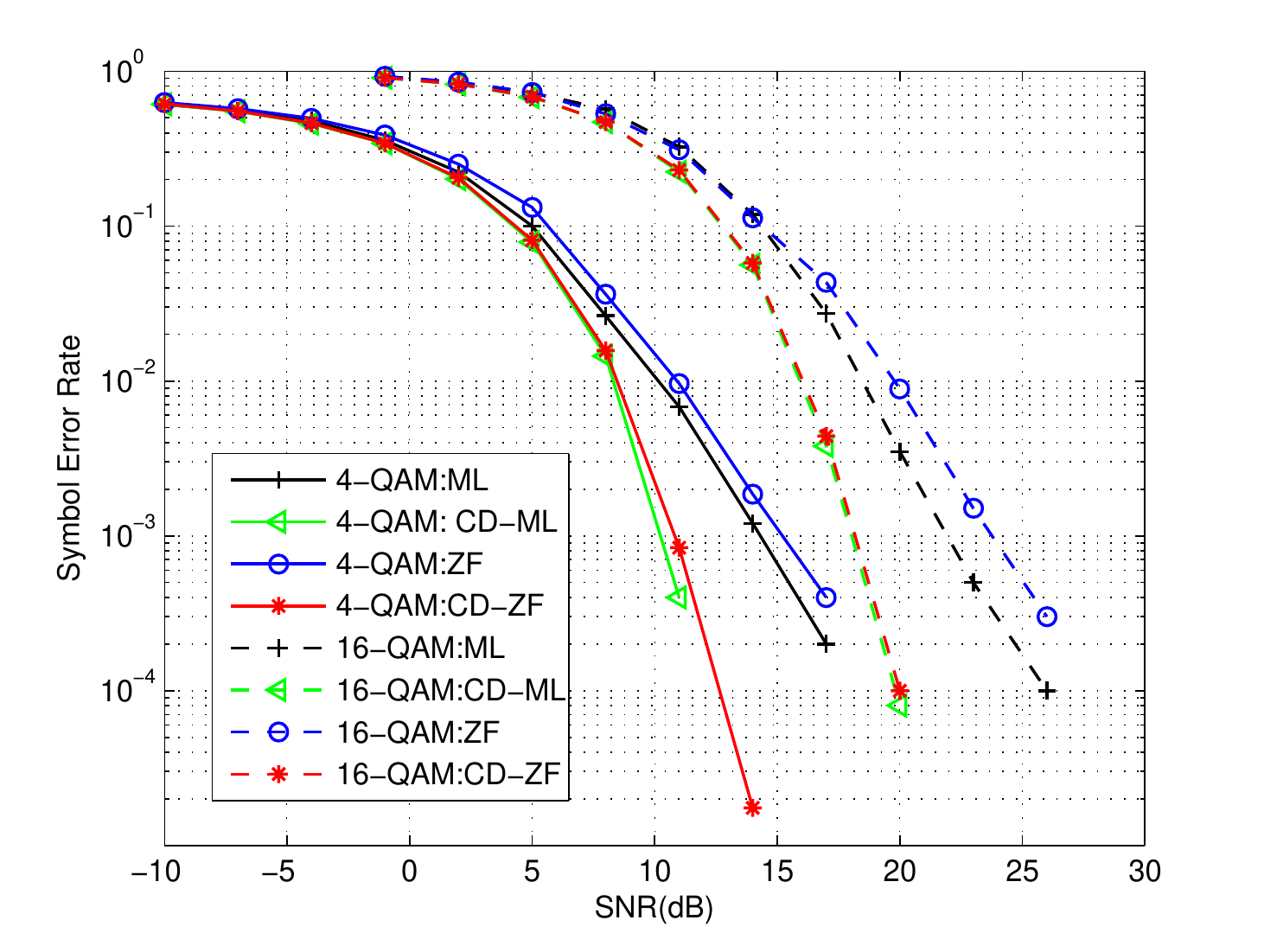}}
\caption{Improvement through Code Diversity for QOSTBC.}
\label{fig:graph1}
\end{center}
\end{figure}

Fig. (\ref{fig:graph1}) shows the performance improvement through
code diversity with two feedback bits (\ie\;\ $K=4$ for the
algorithm described in (\ref{QOSTBCalg2})) using both the ML scheme
(\ref{MLdec}) and the ZF scheme (\ref{zfdec}) for both 4-QAM and
16-QAM. In the Fig. (\ref{fig:graph1}), the prefix 'CD-' denotes
that the corresponding decoding scheme is aided by our code
diversity method.

The property of quasi-orthogonality implies that the ML scheme has
quadratic complexity and the ZF has linear complexity.Two feedback
bits improve performance of ML decoding by about $3$ dB
for both 4-QAM and 16-QAM. The performance loss from suboptimal zero
forcing decoding is minimal.

\textbf{Remark}: Jafarkhani \cite{Jafark2} and Su {\it et al.}
\cite{Su} observed that it is possible to achieve full diversity by
rotating the signal constellation appropriately. However, the
benefits of constellation rotation are limited to small
constellations whereas the benefits of code diversity are not. The method
of inducing channel phases enlarges the code diversity framework introduced by
Tan and Calderbank \cite{tan}.

\subsection{Multi-User Detection of Alamouti Signals}
\label{AlaP}

Consider a two-user system where each user employs the Alamouti
code. The received signals are
\begin{equation}
\begin{array}{ll}
\label{polarizationAM} r_{1}=\mathcal{H}_{1}c_{1}+\mathcal{G}_{1}c_{2}+n_{1}\\
r_{2}=\mathcal{H}_{2}c_{1}+\mathcal{G}_{2}c_{2}+n_{2}
\end{array}
\end{equation}
where $c_{1}=(x_{1},x_{2})^{T}, c_{2}=(x_{3},x_{4})^{T}$ are the
signals transmitted by the first and second users respectively. The
matrices
\begin{equation}
\begin{array}{ll}
\mathcal{H}_{1}=\left (
\begin{array}{cc}
h_{11} & h_{21}\\
-h_{21}^{*} & h_{11}^{*} \end{array} \right),\mathcal{G}_{1}=\left (
\begin{array}{cc}
g_{11} & g_{21}\\
-g_{21}^{*} & g_{11}^{*} \end{array} \right)\\
\\ \mathcal{H}_{2}=\left (
\begin{array}{cc}
h_{12} & h_{22}\\
-h_{22}^{*} & h_{12}^{*} \end{array} \right), \mathcal{G}_{2}=\left
(
\begin{array}{cc}
g_{12} & g_{22}\\
-g_{22}^{*} & g_{12}^{*} \end{array} \right)
\end{array}
\end{equation} are Alamouti blocks, and
 $H\mathcal{}_{2}, \mathcal{G}_{1}$ represent the interference channels.

We can rewrite (\ref{polarizationAM}) as
\begin{equation}
\label{polarizationAM1} \left ( \begin{array}{c}
r_{1}\\
r_{2} \end{array} \right)=\left ( \begin{array}{cc} \mathcal{H} &
\mathcal{G}
\end{array} \right) \left ( \begin{array}{c}
c_{1}\\
c_{2} \end{array} \right) + \left ( \begin{array}{c}
n_{1}\\
n_{2} \end{array} \right)
\end{equation}
where $\mathcal{H} =(\mathcal{H}_{1}, \mathcal{H}_{2})^{T},\;\
\mathcal{G}=(\mathcal{G}_{1}, \mathcal{G}_{2})^{T}$.

Several decoding schemes including ML, Bayesian interference
cancellation and ZF are explored by Sirianumpiboon et al.
\cite{Songsri} where it is shown that
\begin{equation}
\lambda =
\frac{\mathcal{H}^{\dag}\mathcal{G}}{\|\mathcal{H}\|\|\mathcal{G}\|}
\end{equation}
determines decoding performance.

Take the ZF scheme \cite{Naguib} for example. Setting $W = \left (
\begin{array}{cc}
I_{2} & -\mathcal{G}_{1}\mathcal{G}_{2}^{-1}\\
-\mathcal{H}_{2}\mathcal{H}_{1}^{-1} & I_{2} \end{array} \right)$,
we have
\begin{equation}
\label{polarizationAMZF} W\left ( \begin{array}{c}
r_{1}\\
r_{2} \end{array} \right)=\left ( \begin{array}{cc}
\mathcal{H}' & 0\\
0 & \mathcal{G}' \end{array} \right) \left (
\begin{array}{c}
c_{1}\\
c_{2} \end{array} \right) + \left ( \begin{array}{c}
n_{1}^{'}\\
n_{2}^{'} \end{array} \right)
\end{equation}
where
$\mathcal{H}'=\mathcal{H}_{1}-\mathcal{G}_{1}\mathcal{G}_{2}^{-1}\mathcal{H}_{2},\;\
\mathcal{G}'=\mathcal{G}_{2}-\mathcal{H}_{2}\mathcal{H}_{1}^{-1}\mathcal{G}_{1}$.
Thus, $W$ decorrelates two users enabling separate detection. The
effective $\SNR$ \cite{Songsri} for the ZF scheme is
$$\frac{\|\mathcal{H}\|\|\mathcal{G}\|}{\sigma^{2}}(1-\|\lambda\|^{2})$$
When $\|\lambda\|=0$, there is no interference and the ZF decoding scheme is optimal.

\subsubsection{Algorithm Implementation}
We use phase adaptation to introduce code diversity so that
\begin{equation}
\begin{array}{ll}
\mathcal{H}_{1}=\left (
\begin{array}{cc}
h_{11}\gamma_{1} & h_{21}\\
-h_{21}^{*} & h_{11}^{*}\gamma_{1}^{*} \end{array} \right),
\mathcal{G}_{2}=\left (
\begin{array}{cc}
g_{12}\gamma_{2} & g_{22}\\
-g_{22}^{*}& g_{12}^{*}\gamma_{2}^{*} \end{array} \right)\\
\\
\mathcal{H}_{2}=\left (
\begin{array}{cc}
h_{12}\gamma_{1} & h_{22}\\
-h_{22}^{*} & h_{12}^{*}\gamma_{1}^{*} \end{array} \right),
\mathcal{G}_{1}=\left (
\begin{array}{cc}
g_{11}\gamma_{2} & g_{21}\\
-g_{21}^{*} & g_{11}^{*}\gamma_{2}^{*} \end{array} \right)
\end{array}
\end{equation}
where $\gamma_{1}=e^{i2\pi\frac{k_{1}}{K_{1}}},
\gamma_{2}=e^{i2\pi\frac{k_{2}}{K_{2}}}$; let $K_{1}=K_{2}=4$,
corresponding to four bits of feedback. The selection strategy is
\begin{equation}
\label{APalg2} \{\hat{k_{1}}, \hat{k_{2}}\}= \argmin_{
k_{1}\in\{1,2...,K_{1}\}, k_{2}\in\{1,2...,K_{2}\}}\|\lambda\|
\end{equation}

\subsubsection{Simulation}
The channels are modeled as complex Gaussian. \ie
\begin{equation}
\begin{array}{ll}
h_{11}, h_{21}, g_{12}, g_{22}\sim \mathcal{CN}(0,1)\\
h_{12}, h_{22}, g_{11}, g_{21}\sim  \mathcal{CN}(0,\gamma)
\end{array}
\end{equation}
where the signal to interference ratio $\gamma$ is taken to be $0.5$.

\begin{figure}[h!]
\begin{center}
\resizebox{6.9cm}{!}{\includegraphics{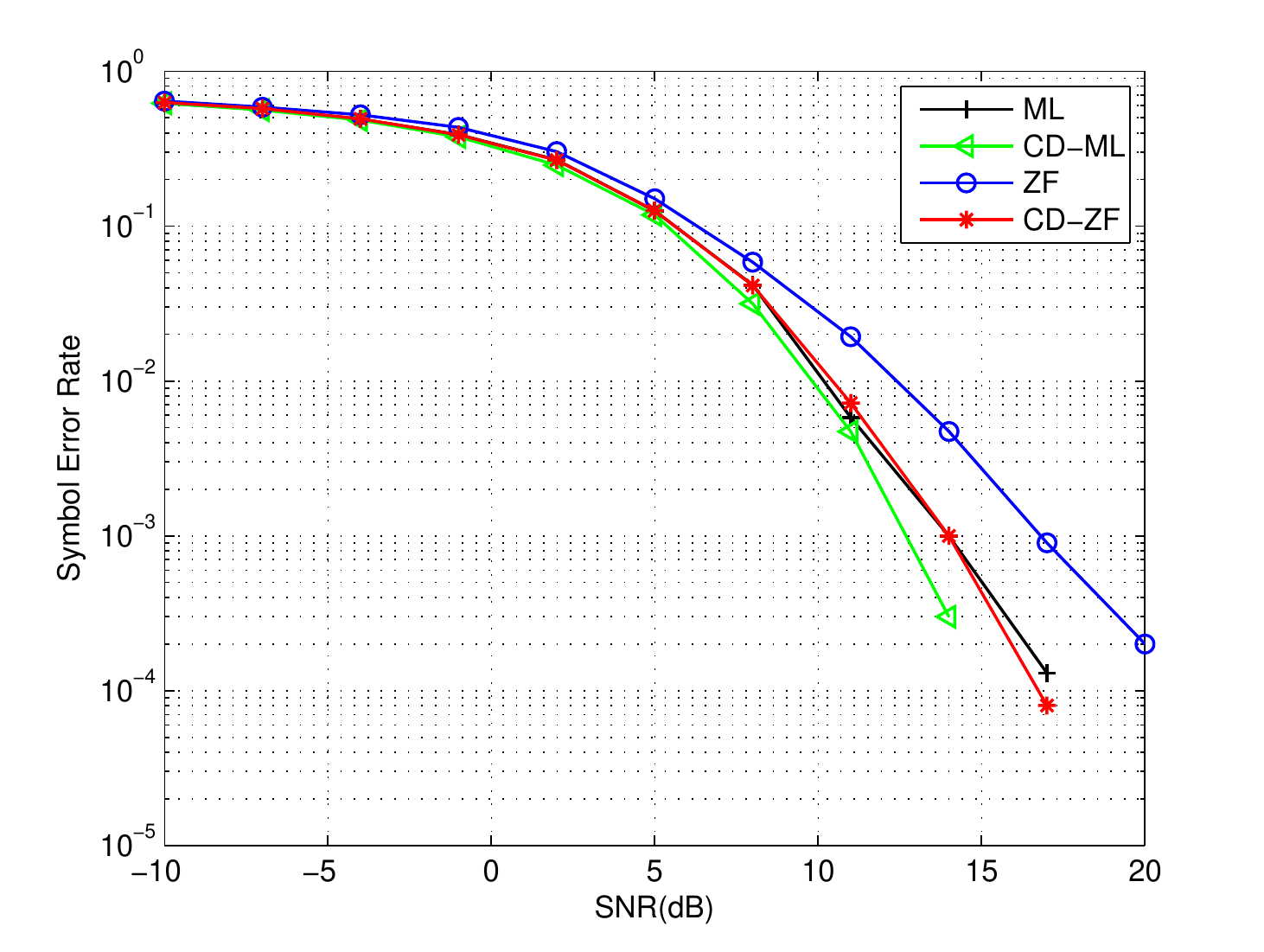}}
\caption{Performance of ZF and ML schemes for Alamouti two-user
detection with code diversity for 4-QAM.} \label{fig:graph4}
\end{center}
\end{figure}

With four bits of feedback \ie\;\ $K_{1}=K_{2}=4$, the performance
of the ML and ZF decoding schemes is shown in Fig. \ref{fig:graph4} where
the prefix 'CD-' indicates that the
corresponding decoding scheme is aided by our code diversity method.

Code diversity improves the performance of ZF decoding by more than
$2$ dB and matches the performance of ML decoding. The performance
of ML decoding is also improved by approximately $1$ dB.

\subsection{Code Diversity for the Golden Code}
Here code diversity is implemented by switching between equivalent variants
of the Golden code.

\subsubsection{Equivalent Variants of Golden Code} \label{recap2} We
recall the structure of the Golden code \cite{Belfiore,Yao,Dayal} as
\begin{equation}
\label{stru0} G_1= \left ( \begin{array}{cc}
s_{1}+\tau s_{2} & s_{3}+\tau s_{4}\\
i(s_{3}+\mu s_{4}) & s_{1}+\mu s_{2} \end{array} \right)
\end{equation}
where $\tau=\frac{1+\sqrt{5}}{2}$ and $\mu=\frac{1-\sqrt{5}}{2}$.

Switching the positions of $\tau$ and $\mu$ yields the variant
\begin{equation}
\label{stru1} G_2= \left ( \begin{array}{cc}
s_{1}+\mu s_{2} & s_{3}+ \mu s_{4}\\
i(s_{3}+\tau s_{4}) & s_{1}+\tau s_{2} \end{array} \right)
\end{equation}

\subsubsection{$2\times 1$ systems}
Let $h_1$ and $h_2$ be channel gains from the two transmit antennas
to the single receive antenna which are modeled as Rayleigh $h_i\sim
\mathcal{CN}(0,\sigma^2),i=1,2$. The noise is complex gaussian as
$\mathcal{CN}(0,N_0)$ and $\mathsf{SNR}$ is defined as
\begin{equation}
\label{SNR} \mathsf{SNR}=\frac{E_s \sigma^2\left
(2+\tau^2+\mu^2\right)}{N_0}
\end{equation}
where $E_s$ is the average transmitting signal power.

Given $h_1$ and $h_2$, the instantaneous $\mathsf{SNR}$ for code
$G_1$ is
$$S_1 = \frac{E_s\{|h_1|^2
(1+\tau^2)+|h_2|^2(1+\mu^2)\}}{N_0}$$

and the instantaneous $\mathsf{SNR}$ for code $G_2$ is
$$S_2 = \frac{E_s\{|h_1|^2
(1+\mu^2)+|h_2|^2(1+\tau^2)\}}{N_0}.$$

We observe $S_1 > S_2$ if and only if $|h_1|>|h_2|$.

Our code diversity scheme is to select $G_1$
if $S_1>S_2$ otherwise to select $G_2$. Code diversity improves the
average $\mathsf{SNR}$ to
\begin{equation}
\label{ASNR} S_{CD}=\frac{E_s\{\mathbf{E}[h_{max}^2]
(1+\mu^2)+\mathbf{E}[h_{min}^2](1+\tau^2)\}}{N_0}
\end{equation}
where
$h_{max}=\max\{|h_1|,|h_2|\},h_{min}=\min\{|h_1|,|h_2|\}$.\vspace{3mm}
\\
\textbf{Proposition 1}: Given Rayleigh channel gains ($f_{|h_i|}(x)=
\frac{x}{\sigma^2}e^{-\frac{x^2}{2\sigma^2}},i=1,2$), the above form
of code diversity improves the performance of ML decoding of the Golden code
by $0.88$ dB.\vspace{2mm}
\\
\textbf{Proof}:
The distribution function for $|h_i|$ is $f_{|h_i|}(x)=
\frac{x}{\sigma^2}e^{-\frac{x^2}{2\sigma^2}},i=1,2$ and the
accumulative distribution function is $F_{|h_i|}(x)=
1-e^{-\frac{x^2}{2\sigma^2}}$. Therefore the distribution function of
$h_{max}=\max\{|h_1|,|h_2|\}$ is given by
\begin{equation}
\begin{array}{ll}
f_{max}(x)=\frac{d\{F_{|h_1|}(x)F_{|h_2|}(x)\}}{dx}\\
\hspace{13.5mm}=\frac{dF^2_{|h_i|}(x)}{dx}=\frac{2x}{\sigma^2}\left(e^{-\frac{x^2}{2\sigma^2}}-e^{-\frac{x^2}{\sigma^2}}\right),
\end{array}
\end{equation}
hence $\mathbf{E}[h_{max}^2]=1.5\sigma^2$.

The distribution function of $h_{min}=\min\{|h_1|,|h_2|\}$ is given by
\begin{equation}
\begin{array}{ll}
f_{min}(x)=\frac{d\left(1-F_{|h_1|}(x)\right)\left(1-F_{|h_2|}(x)\right)}{dx}\\
\hspace{12.5mm} =\frac{2x}{\sigma^2}e^{-\frac{x^2}{\sigma^2}},
\end{array}
\end{equation}

hence $\mathbf{E}[h_{min}^2]=0.5\sigma^2$.

The gain is given by
\begin{equation}
\begin{array}{ll}
G_{1rx}=10\log(\frac{S_{CD}}{\mathsf{SNR}})\\
=10\log\frac{\mathbb{E}[h_{max}^2]
(1+\mu^2)+\mathbb{E}[h_{min}^2](1+\tau^2)\}}{\sigma^2\left
(2+\tau^2+\mu^2\right)}=0.88\;\ dB
\end{array}
\end{equation}
$\square$
\subsubsection{$2\times 2$ systems}
Let $h_1, h_2$ be channel gains from the two transmit antennas to
the first receive antenna and $h_3,h_4$ for the second receive
antenna. The $\mathsf{SNR}$ is the same as defined in (\ref{SNR}).

For $G_1$, the instantaneous $\mathsf{SNR}$ is
$$S_1 = \frac{E_s\{(|h_1|^2+|h_3|^2)
(1+\tau^2)+(|h_2|^2+|h_4|^2)(1+\mu^2)\}}{2N_0}$$

For $G_2$, the instantaneous $\mathsf{SNR}$ is
$$S_2 = \frac{E_s\{(|h_1|^2+|h_3|^2)
(1+\mu^2)+(|h_2|^2+|h_4|^2)(1+\tau^2)\}}{2N_0}$$ and we observe
$$S_1 > S_2
\iff |h_1|^2+|h_3|^2>|h_2|^2+|h_4|^2.$$

Code diversity is implemented by selecting $G_1$ if $S_1>S_2$ otherwise
selecting $G_2$. Monte Carlo simulation shows that under code
diversity, the gain over ML decoding error performance is
\begin{equation}
G_{2rx}=0.66\;\ dB
\end{equation}

\subsubsection{Simulations} 
\begin{figure}[h!]
\begin{center}
\resizebox{7cm}{!}{\includegraphics{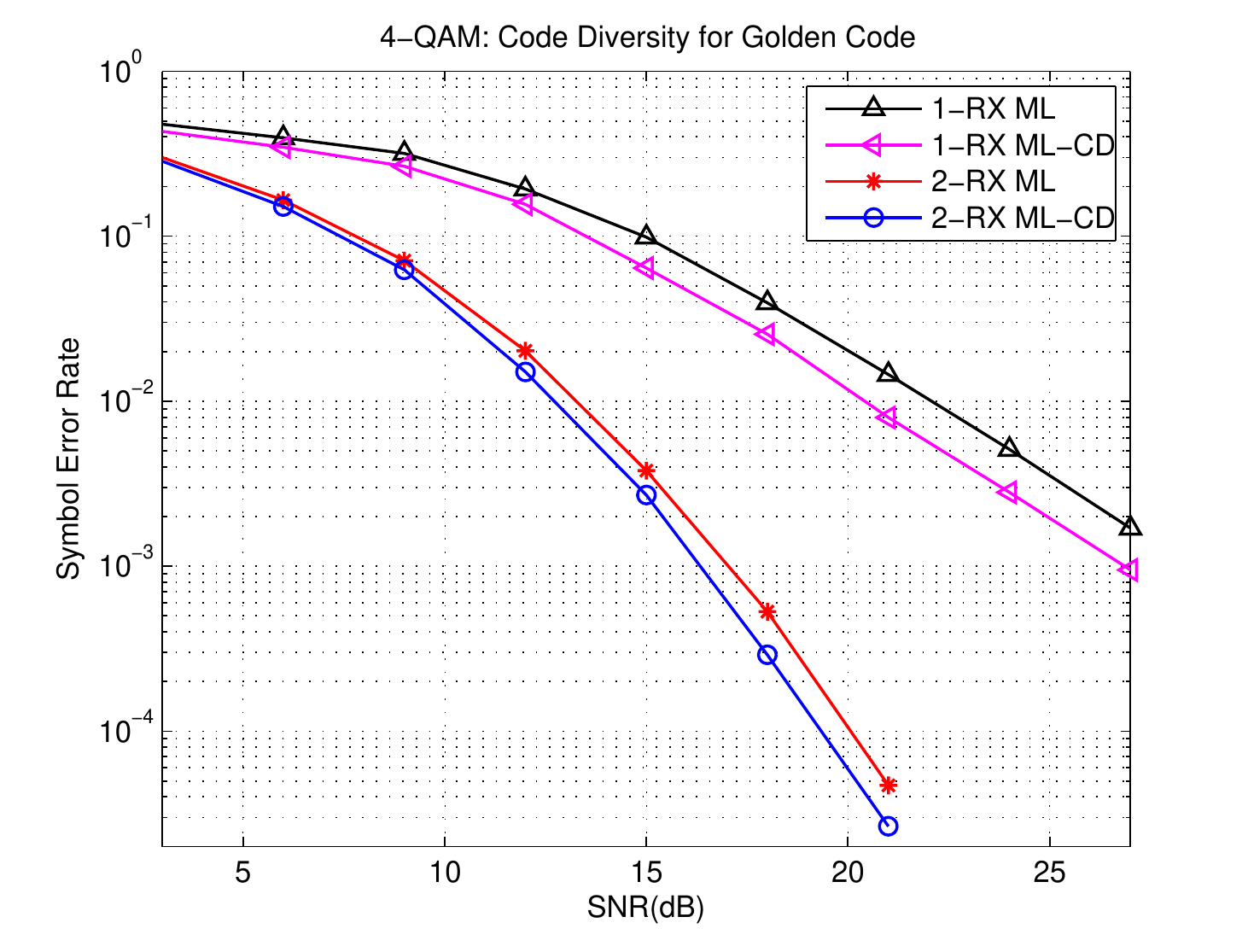}}
\caption{Improvement to the ML Decoder for the Golden code through
Code Diversity.} \label{fig:graph5}
\end{center}
\end{figure}

The simulations shown in Fig. \ref{fig:graph5}
employ a single feedback bit. Performance is improved by about $1$ dB for $2\times 1$ systems and about
$0.5$ dB for $2\times 2$ systems; the simulation result is
consistent with the analysis.

\section{A family of Full Rate Circulant Code Designs based on Code Diversity}
\label{CCD} Observing the rarity of full rate $M\times M$ code
designs as $M$ becomes relatively large, we propose a family of full
rate codes based on circulant matrix in this section. Code designs
based on circulant matrix without feedback are studied. We also
illustrate a universal linear decoder using Fourier basis for the
family of circulant codes within the code diversity framework.

\subsection{Circulant Matrices and Their Properties}
An $M\times M$ circulant matrix takes the form
\begin{equation}
\label{CirculantCode}  X=\left ( \begin{array}{c}
\vec{x}\\
\vec{x}L\\
\cdots\\
\vec{x}L^{M-1}\end{array} \right)
\end{equation}
where $\vec{x}=(x_1,x_2,\cdots,x_M)$ and $(x_1,x_2,\cdots,x_M)L=(x_M,x_1,\cdots,x_{M-1})$. We recall the following properties of circulant matrices and refer the reader to \cite{Davis} for more details.\vspace{3mm}
\\
\textbf{Lemma 1}: Let $\omega=e^{\frac{2\pi i}{M}}$. The eigenvectors of the shift operator L are the Fourier basis
$f_{j}=(1, \omega^j, \omega^{2j},\cdots, \omega^{(M-1)j})^T,j=0,1,\cdots,M-1$ and $L f_j=\omega^j f_j$.\vspace{3mm}
\\
\textbf{Lemma 2}: A matrix $X$ commutes with the left shift matrix
$L$ if and only if $X$ is a circulant matrix.\vspace{3mm}
\\
\textbf{Property 1}: Let $X$ be an $M\times M$ circulant design. Then
the Fourier basis $F=\{f_0,\cdots,f_{M-1}\}$ is an orthogonal set of
eigenvectors of $X$.\vspace{2mm}
\\
\textbf{Proof}: We have $L(Xf_j)=XL f_j =\lambda_{j}(L) X f_j, j=0,1,\cdots,M-1$ and so $X f_j$ is a multiple of $f_j$.$\square$\vspace{3mm}
\\
\textbf{Property 2}: Let $X$ be an $M\times M$ circulant design with first row
$\vec{x}$. Then the $j^{th}$ eigenvalue of
$X$ is $\lambda_{j}(X)=\vec{x} f_j$.\vspace{2mm}
\\
\textbf{Proof}:
\begin{equation}
\begin{array}{ll}
X f_j=\left ( \begin{array}{c}
\vec{x}\\
\vec{x}L\\
\cdots\\
\vec{x}L^{M-1}\end{array} \right) f_j = \left ( \begin{array}{c}
\vec{x}f_j\\
\vec{x}Lf_j\\
\cdots\\
\vec{x}L^{M-1}f_j\end{array} \right)\\
\\
=\vec{x}f_j \left (
\begin{array}{c}
1\\
\omega^j\\
\cdots\\
\omega^{(M-1)j}\end{array} \right) =(\vec{x}f_j)f_j
\end{array}
\end{equation}
therefore, $\lambda_{j}(X)=\vec{x} f_j$. $\square$

\subsection{$3\times3$ Circulant Design without Feedback}
Consider the $3\times3$ circulant code $C$ given by
\begin{equation}
\label{CCode}  C=\left ( \begin{array}{ccc}
x_{1}\alpha & x_{2}\beta & x_{3}\\
x_{2}\beta & x_{3} & x_{1}\alpha\\
x_{3} & x_{1}\alpha & x_{2}\beta \end{array} \right)
\end{equation}
where $\alpha=(\frac{1+\sqrt{5}}{2})^\frac{1}{3}$,
$\beta=(\frac{1-\sqrt{5}}{2})^\frac{1}{3}$ and $x_i$ are symbols
from QAM constellation.

\subsubsection{Full Rate}
This $3\times3$ circulant design is full rate since it transmits three symbols
over three consecutive time slots.

\subsubsection{Full Diversity}
Consider two distinct codewords $C_1$ and $C_2$ with symbols on
QAM constellation. Then,
$$|det(C_1-C_2)|=\left|\frac{1+\sqrt{5}}{2}s_1^3+\frac{1-\sqrt{5}}{2}s_2^3+s_3^3+3s_1 s_2
s_3\right|$$ where $s_i, i=1,2,3$ are the corresponding symbol differences and
$s_i$ are gaussian integers.
\begin{itemize}
\item if $s_1 \neq s_2$, then $C_1-C_2$ is irrational and so $|det(C_1-C_2)|\neq 0$.
\item if $s_1 =s_2$, then $|det(C_1-C_2)| = 0$ if and only if
$s_1^3+s_3^3+3s_1^2s_3=0$. Since $z^3+3z+1=0$ has no solution in $Q(i)$, so
$s_1^3+s_3^3+3s_1^2s_3=0$ has no solution in $Z(i)$. Therefore, $|det(C_1-C_2)| \neq 0$.
\end{itemize}

Hence, it has full diversity.

\subsubsection{Simulation} 
\begin{figure}[h!]
\begin{center}
\resizebox{7cm}{!}{\includegraphics{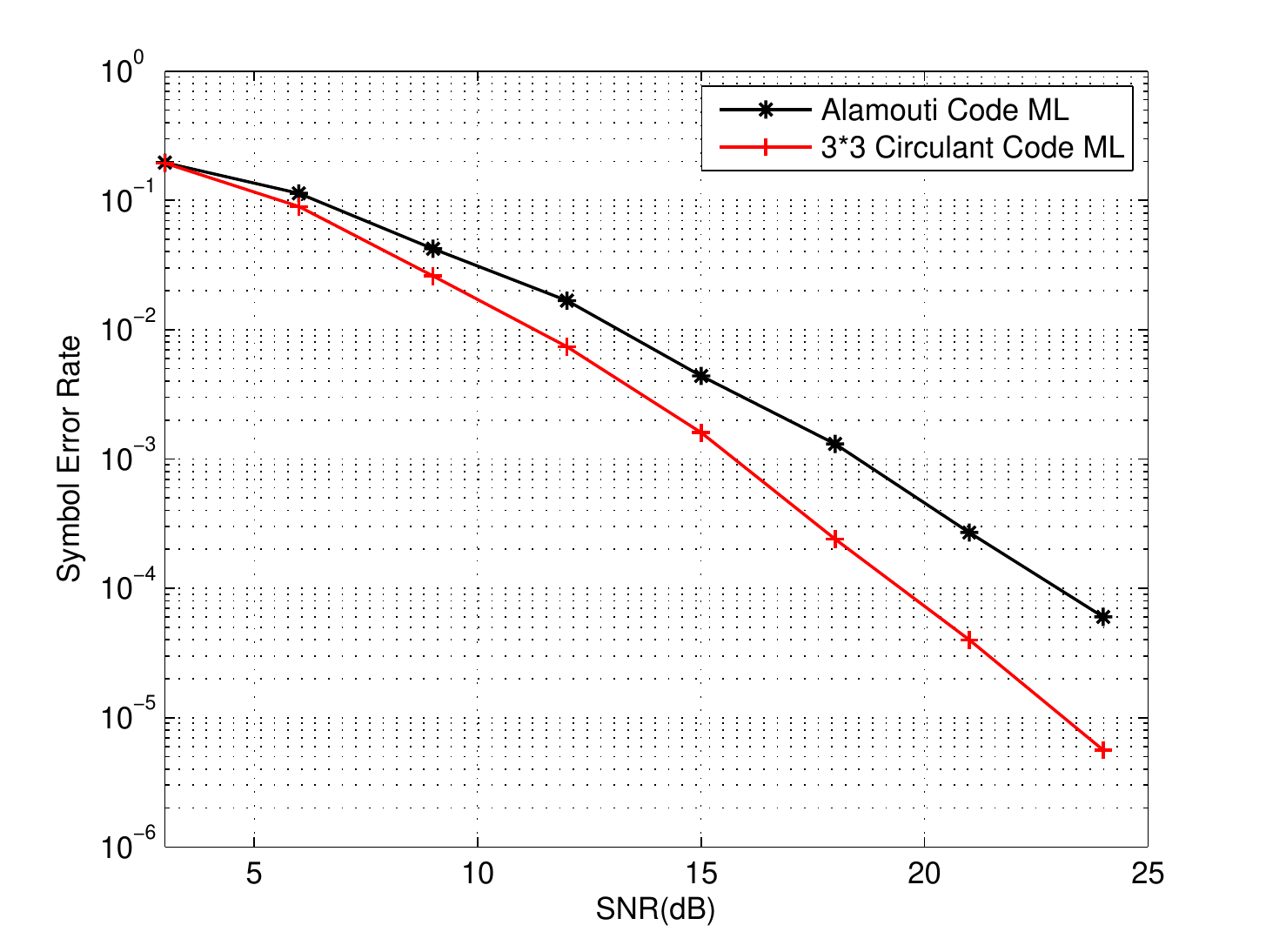}} \caption{Performance comparison of the $3\times
3$ circulant code with the Alamouti code. Curvers are for ML decoding and
$4-$QAM signaling.} \label{fig:graph6}
\end{center}
\end{figure}

We compare this $3\times 3$ circulant code
with the Alamouti code at the same transmission rate of two bits per
time slot in Fig. \ref{fig:graph6} (both using 4-QAM signaling). The $3\times 3$
circulant code outperforms the Alamouti code significantly and has a larger
diversity which is indicated by the slope of the curves.

\textbf{Remark}: General $M\times M$ circulant codes can be produced using similar techniques.

\subsection{Integration of Circulant Codes and Code Diversity}
\label{CCCD} With code diversity, we use circulant matrix in
(\ref{CirculantCode}) or their variants such as the one in
(\ref{CCode}) as our coding block. We observe that the circulant structure transfers from the
transmitter to the receiver (of (\ref{model1})), in that the received signal $r$ can be written as
\begin{equation}
\label{GCC} r=\mathcal{H}c+n
\end{equation}
where $c=(x_1,x_2,\cdots,x_M)^T$, $x_i$ is taken from certain
constellations, $n$ is the corresponding additive noise and
\begin{equation}
\label{Chcc}  \mathcal{H}=\left ( \begin{array}{c}
\vec{h}\\
\vec{h}L\\
\cdots\\
\vec{h}L^{M-1}\end{array} \right)
\end{equation}
where $\vec{h}=(h_1,h_2,\cdots,h_M)$ with $h_j$ as the channel gain
between the $j^{th}$ transmit antenna and the single receive
antenna.

\subsubsection{Code Diversity Algorithm}
We modify the phase of the channel gain $h_1$ as follows:
\begin{equation}
\begin{array}{ll}
\hat{k} =\argmax_{k\in K} det(\mathcal{H}^{\dag}\mathcal{H})
|_{h_1\rightarrow
h_1 e^{\frac{i k \pi}{K} }}\\
\\
\hspace{5mm}= \argmax_{k\in
K}\left|\prod^{N}_{j=1}\vec{h}f_j\right|_{h_1\rightarrow h_1
e^{\frac{i k \pi}{K} }}
\end{array}
\end{equation}
where $\log_2 K$ is the number of feedback bits. More channel gains may need
to be modified as $M$ increases.

\subsubsection{Linear decoder based on Fourier Basis}
From equation (\ref{GCC}), we have
\begin{equation}
F^{-1}r = F^{-1}\mathcal{H}FF^{-1}c + F^{-1}n
\end{equation}
and we observe that $F^{-1}=\frac{1}{M} F'$, where $F'$ is the
entry-wise conjugate of $F$, so we get
\begin{equation}
F' r = \Lambda_{\mathcal{H}}F' c + F' n
\end{equation}
where $\Lambda_{\mathcal{H}} = F^{-1}\mathcal{H}F = \mathsf{diag}
\{\vec{h}f_0,\cdots,\vec{h}f_{M-1}\}$. We also observe that $f_j^T
F'c  = M x_j, \forall j=1,\cdots,M$.

We propose the linear decoder as
\begin{equation}
\label{LinearA} \hat{x}_j = \argmin_{x_j \in
\mathbb{C}}\left|\frac{1}{M}f_j^T \Lambda^{-1}_{\mathcal{H}} F'
c-x_j\right|^2
\end{equation}

The decoding performance of the linear decoder mainly depends on
$\det{\Lambda_{\mathcal{H}}}$ which is exactly
$(\det{\mathcal{H}^{\dag}\mathcal{H}})^{\frac{1}{2}}$. Our code
diversity algorithm precisely aims to raise
$\det{\mathcal{H}^{\dag}\mathcal{H}}$. Therefore, the performance of
the linear decoder is guaranteed within the code diversity framework.

%\subsubsection{Simulations} In our simulation, we take the $3\times
%3$ circulant code (\ref{CCode}) as an example. We use two feedback
%bits for the linear decoding of the $3\times 3$ circulant code.
%\begin{figure}[h!]
%\begin{center}
%\resizebox{7cm}{!}{\includegraphics{CC33CD_M.eps}} \caption{Linear
%decoding for $3\times 3$ circulant code through code diversity}
%\label{fig:graph7}
%\end{center}
%\end{figure}

%In the Fig. \ref{fig:graph7}, '$3*3$ circulant code-CD LD' denotes
%the curve for the linear decoding of $3\times 3$ circulant code
%through code diversity. It shows that linear decoding with two
%feedback bits obtains close optimal performance.

\section{Conclusion}
\label{conclusion} We have introduced a general information
theoretic framework for the analysis of code diversity. we have
shown that it not only improves the diversity and coding advantages
for general space time codes but enables optimal decoding
performance with low complexity decoding and only a small number of
feedback bits. The method of code diversity also reduces the
capacity loss associated with some forms of space-time coding. We
have also proposed a family of full rate circulant code designs
based on code diversity and a corresponding universal linear
decoding algorithm using decomposition of circulant matrices. An
extensive study of $3\times 3$ circulant design shows that it
outperforms the Alamouti code at the same transmission rate.

\section{Acknowledgement}
\label{acknowledgement} The authors would like to thank Chee Wei Tan
for many helpful suggestions and for bringing the work of Bonnet
{\it et al.} to their attention.

\end{document}